\documentclass[aps,pra,english,twocolumn,notitlepage,showpacs,nofootinbib]{revtex4-1}
\usepackage{mathptmx}
\usepackage[T1]{fontenc}
\usepackage[latin9]{inputenc}
\usepackage{color}
\usepackage{babel}
\usepackage{textcomp}
\usepackage{bm}
\usepackage{graphicx}
\usepackage{amsmath}
\usepackage{amsthm}
\usepackage{amssymb}
\usepackage{epstopdf}
\usepackage[unicode=true,pdfusetitle,
bookmarks=true,bookmarksnumbered=false,bookmarksopen=false,
breaklinks=true,pdfborder={0 0 0},pdfborderstyle={},backref=false,colorlinks=true]{hyperref}
\usepackage{amsmath, amsfonts, amssymb, amsthm, mathrsfs, amssymb, braket, bbm, bm,graphicx,times,txfonts,color,subfigure}
\usepackage{hyperref}
\usepackage[table]{xcolor}

\hypersetup{
urlcolor=blue, citecolor=blue, linkcolor=blue}
\makeatletter
\theoremstyle{plain}
\newtheorem{Theorem}{Theorem}
\newtheorem{Proposition}{Proposition}

\newtheorem{Lemma}{Lemma}

\makeatother
\begin{document}

\title{Numerical ranges of Bargmann invariants}
\author{Jianwei Xu}
\email{xxujianwei@163.com}

\begin{abstract}
Bargmann invariants have recently emerged as powerful tools in quantum information theory. Despite their widespread use, a complete characterization of their physically realizable values has remained an outstanding challenge. In this work, we provide a rigorous determination of the numerical range of Bargmann invariants for quantum systems of arbitrary finite dimension. We demonstrate that any permissible value of these invariants can be achieved using either (i) pure states exhibiting circular Gram matrix symmetry or (ii) qubit states alone. These results establish fundamental limits on Bargmann invariants in quantum mechanics and provide a solid mathematical foundation for their diverse applications in quantum information processing.
\end{abstract}

\maketitle

%\affiliation{College of Science, Northwest A$\&$F University, Yangling, Shaanxi 712100, China}

% \date{\today }
% \pacs{03.65.Ud, 03.67.Mn, 03.65.Aa}
% \date{\today }

\section{Introduction}
%\setcounter{equation}{0} \renewcommand%
%\theequation{1.\arabic{equation}}

In quantum information theory, the analysis of $n$-tuples of quantum states $\Gamma = (\rho_1, \rho_2, \dots, \rho_n)$, where $\{\rho_j\}_{j=1}^n$ are density operators on a $d$-dimensional complex Hilbert space $\mathbb{C}^d$, plays a fundamental role in numerous applications.  A natural question is how to describe the
properties of $\Gamma =\left( \rho _{1},\rho _{2},...,\rho _{n}\right) .$
For example, does there exist a unitary $U$ such that $U\Gamma U^{\dagger
}=\left( U\rho _{1}U^{\dagger },U\rho _{2}U^{\dagger },...,U\rho
_{n}U^{\dagger }\right) $ becomes a tuple of diagonal states \cite{Brunner-2021-PRL}, or
becomes a tuple of real states \cite{Miyazaki-2022-Quantum,Fernandes-2024-PRL}? Where we use $%
U^{\dagger }$ to denote the conjugate
transpose of $U.$ One property of $\Gamma =\left(
\rho _{1},\rho _{2},...,\rho _{n}\right) $ is its $n$-order Bargmann
invariant or multivariate trace \cite{Bargmann-1964-JMP,Oszmaniec-2024-NJP,Wilde-2024-Quantum}  Tr$\left( \rho _{1}\rho _{2}...\rho _{n}\right) .$ One basic
property of Tr$\left( \rho _{1}\rho _{2}...\rho _{n}\right) $ is that it is
invariant under the unitary action $U\Gamma U^{\dagger }=\left( U\rho
_{1}U^{\dagger },U\rho _{2}U^{\dagger },...,U\rho _{n}U^{\dagger }\right) .$

Bargmann invariants have diverse applications in quantum information theory, such as in geometric phases \cite{Simon-1993-PRL}, photonic indistinguishability \cite{Menssen-2017-PRL,Jones-2020-PRL}, quantum error mitigation \cite{Fei-2023-PRA}, scalar spin chirality \cite{Galvao-2023-PRR}, and weak values \cite{Wagner-2023-PRA}. Every element of Kirkwood-Dirac distribution is just a 3-order Bargmann invariant \cite{Kirkwood-1933-PR,Dirac-1945-RMP,Bievre-2024-NJP}.

The numerical range of Tr$\left( \rho _{1}\rho _{2}...\rho _{n}\right) $ for
all $\Gamma =\left( \rho _{1},\rho _{2},...,\rho _{n}\right) $ is
defined as the set
\begin{equation}
\boldsymbol{B}_{n,d}=\left\{ \text{Tr}\left( \rho _{1}\rho _{2}...\rho
_{n}\right) :\{ \rho _{j}\} _{j=1}^{n}\subseteq \mathbf{D}%
_{d}\right\}             \label{eq1-1}
\end{equation}%
with $\mathbf{D}_{d}$ the set of all density operators on $\mathbb{C}^{d},$ we also use $\mathbf{D}_{d}^{n}$ to denote the set of all $\Gamma
=\left( \rho _{1},\rho _{2},...,\rho _{n}\right) .$ Clearly, $\boldsymbol{B}%
_{n,d}\subseteq\mathbb{C}$ with $\mathbb{C}$ the set of complex numbers. When the states in Eq. (\ref{eq1-1}) are all (normalized) pure states, $\Psi =\left( |\psi _{1}\rangle ,|\psi _{2}\rangle
,...,|\psi _{n}\rangle \right) ,$ we define
\begin{equation}
\mathcal{B}_{n,d}=\left\{ \text{Tr}\left( |\psi _{1}\rangle \langle \psi
_{1}|\psi _{2}\rangle \langle \psi _{2}|...|\psi _{n}\rangle \langle \psi
_{n}|\right) :\{ |\psi _{j}\rangle \} _{j=1}^{n}\subseteq \mathbf{%
P}_{d}\right\}             \label{eq1-2}
\end{equation}%
with $\mathbf{P}_{d}$ the set of all normalized vectors in $\mathbb{C}^{d}$, we also use $\mathbf{P}_{d}^{n}$ to denote the set of all $\Psi
=\left( |\psi _{1}\rangle ,|\psi _{2}\rangle ,...,|\psi _{n}\rangle \right) .
$ Clearly,
\begin{equation}
\mathcal{B}_{n,d}\subseteq \boldsymbol{B}_{n,d}.           \label{eq1-3}
\end{equation}

For $z\in \mathcal{B}_{n,d},$ by the definition of $\mathcal{B}_{n,d}$ in Eq.
(\ref{eq1-2}), there exists a tuple $\Psi =\left( |\psi _{1}\rangle ,|\psi _{2}\rangle
,...,|\psi _{n}\rangle \right) \in \mathbf{P}_{d}^{n}$ such that $z=$Tr$%
\left( |\psi _{1}\rangle \langle \psi _{1}|...|\psi _{n}\rangle \langle \psi
_{n}|\right) ,$ we say that the value $z$ is quantum realizable by the tuple
$\Psi =\left( |\psi _{1}\rangle ,|\psi _{2}\rangle ,...,|\psi _{n}\rangle
\right) .$ Similarly, for $z\in \boldsymbol{B}_{n,d},$ by the definition of $%
\boldsymbol{B}_{n,d}$ in Eq. (\ref{eq1-1}), there exists a tuple $\Gamma =\left( \rho
_{1},\rho _{2},...,\rho _{n}\right) \in \mathbf{D}_{d}^{n}$ such that $z=$Tr$%
\left( \rho _{1}\rho _{2}...\rho _{n}\right) ,$ we say that the value $z$ is
quantum realizable by the tuple $\Gamma =\left( \rho _{1},\rho _{2},...,\rho
_{n}\right) .$

Some properties of $\Gamma =\left( \rho _{1},\rho _{2},...,\rho _{n}\right) $
are related to $B_{n,d}$, see Refs. \cite{Fernandes-2024-PRL,Oszmaniec-2024-NJP}. Some
results of $\boldsymbol{B}_{n,d}$ and $\mathcal{B}_{n,d}$ have been obtained
in Refs. \cite{Fernandes-2024-PRL,Oszmaniec-2024-NJP,Li-2025-PRA,Zhang-2025-PRA}, but the sets $\boldsymbol{B}_{n,d}$ and $\mathcal{B}_{n,d}$
have not been completely determined till now. In this work, we completely
determine the sets of $\boldsymbol{B}_{n,d}$ and $\mathcal{B}_{n,d},$ i.e.,
the numerical ranges of Bargmann invariants.

This paper is structured as follows. In \hyperlink{section II}{section II}, we provide some
preliminaries and review some existing results about $\boldsymbol{B}_{n,d}$
and $\mathcal{B}_{n,d}.$ In \hyperlink{section III}{section III}, we completely determine $%
\boldsymbol{B}_{n,d}$ and $\mathcal{B}_{n,d}.$ \hyperlink{section IV}{Section IV} is a brief summary.

\hypertarget{section II}{}
\section{preliminaries and some existing results}
%\setcounter{equation}{0} \renewcommand%
%\theequation{2.\arabic{equation}}

In this section, we introduce the notation we will use, provide some
preliminaries and review some existing results about $\boldsymbol{B}_{n,d}$
and $\mathcal{B}_{n,d}.$

\subsection{Notation and basic facts}

Let $\mathbb{Z},$ $\mathbb{R},$ $\mathbb{R}^{+}$ denote the sets of integers, real numbers and positive real numbers. $i=\sqrt{-1}$ is the imaginary unit. We use $|z|$ to denote the absolute value of a
complex number or real number $z.$  Let $\llbracket{j,k}\rrbracket$ stand for the set of
integers from $j$ to $k,$ for example, $\llbracket{2,5}\rrbracket=\{2,3,4,5\}.$ For an $%
n\times n$ matrix $H,$ the entries $\{H_{jk}\}_{j,k=1}^{n}$ can be allowed
to use the integer indices  beyond $\llbracket{1,n}\rrbracket$ by taking the modulus (mod $n$)
automatically, for example, $H_{0,-1}=H_{n,n-1}.$ Similarly, for the triple $%
\Psi =\left( |\psi _{1}\rangle ,...,|\psi _{n}\rangle \right) ,$ we use $%
|\psi _{n}\rangle =|\psi _{0}\rangle $ and $|\psi _{n+1}\rangle =|\psi
_{1}\rangle $ etc.

For any tuple $\Gamma =\left( \rho _{1},\rho _{2},...,\rho _{n}\right) $ of
states $\{\rho _{j}\}_{j=1}^{n}$, by the eigendecompositions of $\rho
_{j}=\Sigma _{\alpha _{j}=1}^{d}\lambda _{j,\alpha _{j}}|\varphi _{j,\alpha
_{j}}\rangle \langle \varphi _{j,\alpha _{j}}|,$ we have
\begin{eqnarray}
&&\text{Tr}\left( \rho _{1}\rho _{2}...\rho _{n}\right)    \nonumber\\
&=&\Sigma _{\alpha _{1},...,\alpha _{n}=1}^{d}\lambda _{1,\alpha
_{1}}...\lambda _{n,\alpha _{n}}\text{Tr}(|\varphi _{1,\alpha _{1}}\rangle
\langle \varphi _{1,\alpha _{1}}|...|\varphi _{n,\alpha _{n}}\rangle \langle
\varphi _{n,\alpha _{n}}|).    \ \ \ \ \ \   \label{eq2-1}
\end{eqnarray}
This shows that%
\begin{equation}
\boldsymbol{B}_{n,d}\subseteq \text{conv}(\mathcal{B}_{n,d}),     \label{eq2-2}
\end{equation}
where conv$(\mathcal{B}_{n,d})$ denotes the convex hull of the set $\mathcal{%
B}_{n,d}.$

For any tuple $\Psi =\left( |\psi _{1}\rangle ,...,|\psi _{n}\rangle \right)
,$ let $|\psi _{n+1}\rangle =|\psi _{n}\rangle ,$ then we have
\begin{eqnarray}
&&\text{Tr}\left( |\psi _{1}\rangle \langle \psi _{1}|...|\psi _{n}\rangle
\langle \psi _{n}|\right)    \notag \\
&=&\text{Tr}\left( |\psi _{1}\rangle \langle \psi _{1}|...|\psi _{n}\rangle
\langle \psi _{n}|\psi _{n+1}\rangle \langle \psi _{n+1}|\right).      \label{eq2-3}
\end{eqnarray}
This says
\begin{equation}
\mathcal{B}_{n,d}\subseteq \mathcal{B}_{n+1,d}.     \label{eq2-4}
\end{equation}

For $\{|\psi _{j}\rangle \}_{j=1}^{n}\subseteq\mathbb{C}^{d_{1}},$ $\{|\psi _{j}\rangle \}_{j=1}^{n}\subseteq\mathbb{C}^{d_{2}},$ $d_{1}\leq d_{2},$ we can always regard $\mathbb{C}^{d_{1}}$ as a subspace of $\mathbb{C}^{d_{2}},$ then $\{|\psi _{j}\rangle \}_{j=1}^{n}\subseteq
\mathbb{C}^{d_{1}}\subseteq\mathbb{C}^{d_{2}}.$ This fact tells that
\begin{equation}
\mathcal{B}_{n,d}\subseteq \mathcal{B}_{n,d+1}.    \label{eq2-5}
\end{equation}

If $n\leq d,$ since dim(span$(\{|\psi _{j}\rangle \}_{j=1}^{n})$)$\leq n,$
thus
\begin{equation}
\mathcal{B}_{n,d}=\mathcal{B}_{n,n}\text{ if }n\leq d.    \label{eq2-6}
\end{equation}

With Eqs. (\ref{eq2-2},\ref{eq2-4},\ref{eq2-5},\ref{eq2-6}), we have
\begin{eqnarray}
\boldsymbol{B}_{n,d} &\subseteq &\boldsymbol{B}_{n+1,d};    \label{eq2-7}  \\
\boldsymbol{B}_{n,d} &\subseteq &\boldsymbol{B}_{n,d+1};    \label{eq2-8}  \\
\boldsymbol{B}_{n,d} &=&\boldsymbol{B}_{n,n}\text{ if }n\leq d.    \label{eq2-9}
\end{eqnarray}
For simplicity, we denote $\mathcal{B}_{n,n}=\mathcal{B}_{n}$ and $%
\boldsymbol{B}_{n,n}=\boldsymbol{B}_{n}.$

We discuss some special cases. If $n=1,$ then $\mathcal{B}_{n,d}=\boldsymbol{%
B}_{n,d}=\{1\}$ for any $d\geq 1.$ If $d=1,$ then $\mathcal{B}_{n,d}=%
\boldsymbol{B}_{n,d}=\{1\}$ for any $n\geq 1.$ When $n=2$ $,$ then $\mathcal{%
B}_{n,d}=\boldsymbol{B}_{n,d}=[0,1]$ for any $d\geq 2.$ With Eqs. (\ref{eq2-4},\ref{eq2-5},\ref{eq2-6},\ref{eq2-7},\ref{eq2-8},\ref{eq2-9}), below,
we mainly consider the cases of $3\leq n\geq d\geq 2.$

\subsection{Gram matrices and circular Gram matrices}

For $\Psi =\left( |\psi _{1}\rangle ,|\psi _{2}\rangle ,...,|\psi
_{n}\rangle \right) \in \mathbf{P}_{d}^{n},$ the Gram matrix (see for example Theorem 7.2.10 in Ref. \cite{Horn-2013-book}) of
$\Psi $ is defined as $G_{\Psi }=((G_{\Psi })_{jk})$ with entries $(G_{\Psi
})_{jk}=\langle \psi _{j}|\psi _{k}\rangle .$ $G_{\Psi }$ is positive
semidefinite, i.e., $G_{\Psi }\succeq 0,$ rank$G_{\Psi }=$dim(span($\{
|\psi _{j}\rangle\} _{j=1}^{n}$)). Conversely, Lemma 1 below holds.

\begin{Lemma} \label{Lemma-1}
(see Ref. \cite{Winter-2004-IJQI}). Let $H$ be an $n\times n$ matrix.
Then, $H$ is positive semidefinite with principal diagonal entries $H_{jj}=1$
if and only if there exists $d\geq 2$ ($d$ depends on $H$) and some $\Psi
=\left( |\psi _{1}\rangle ,|\psi _{2}\rangle ,...,|\psi _{n}\rangle \right)
\in \mathbf{P}_{d}^{n}$ such that $H=G_{\Psi }.$
\end{Lemma}

Notice that the Bargmann invariant
\begin{equation}
\text{Tr}\left( |\psi _{1}\rangle \langle \psi _{1}|...|\psi _{n}\rangle
\langle \psi _{n}|\right) =\Pi _{j=1}^{n}\langle \psi _{j}|\psi
_{j+1}\rangle =\Pi _{j=1}^{n}(G_{\Psi })_{j,j+1}   \ \ \ \  \ \ \ \   \label{eq2-10}
\end{equation}%
is invariant under the choice of global phases, i.e., under the
transformation $\{ |\psi _{j}\rangle \} _{j=1}^{n}\rightarrow
\{ e^{i\theta _{j}}|\psi _{j}\rangle\} _{j=1}^{n},$ here $%
\{\theta _{j}\}_{j=1}^{n}\subseteq\mathbb{R};$ however $(G_{\Psi })_{jk}=\langle \psi _{j}|\psi _{k}\rangle $ is not invariant under the choice of global phases in general.

Let $\mathbf{G}_{n}$ denote the set of all $n\times n$ matrices $H$
satisfying $H\succeq 0$ and $H_{jj}=1$ for $j\in \llbracket{1,n}\rrbracket,$
\begin{equation}
\mathbf{G}_{n}=\{H:H\succeq 0,H_{jj}=1 \ \forall j \in \llbracket{1,n}\rrbracket\}.           \label{eq2-11}
\end{equation}
By the definition of $\mathbf{G}_{n}$ and Lemma \ref{Lemma-1}, $z\in \mathcal{B}_{n}$ if and only if there exists $H\in
\mathbf{G}_{n}$ such that $z=\Pi _{j=1}^{n}H_{j,j+1}.$

$\mathbf{G}_{n}$ has the following properties in Lemma 2.

\begin{Lemma} \label{Lemma-2}  \text{ \ }   \hypertarget{2.a}{}

(2.a). $\mathbf{G}_{n}$ is closed under Hadamard product.  \hypertarget{2.b}{}

(2.b). $\mathbf{G}_{n}$ is closed under convex combination.
\end{Lemma}

(2.a) says that if $H_{1}\in \mathbf{G}_{n},$ $H_{2}\in \mathbf{G}_{n},$ then $H_{1}\circ
H_{2}\in \mathbf{G}_{n}.$ Where $H_{1}\circ H_{2}$ is the Hadamard product of $H_{1}$
and $H_{2},$ i.e., $(H_{1}\circ H_{2})_{jk}=(H_{1})_{jk}(H_{2})_{jk}.$ $%
(H_{1}\circ H_{2})_{jj}=1$ is obvious. $H_{1}\circ H_{2}\succeq 0$ is the
result of Schur product theorem (see for example Theorem 7.5.3 in Ref. \cite{Horn-2013-book}).

(2.b) says that if $\{H_{l}\}_{l=1}^{n}\in \mathbf{G}_{n},$ $\{p_{l}\}_{l=1}^{n}$ is
a probability distribution, i.e., $\Sigma _{l=1}^{n}p_{l}=1,$ $p_{l}\geq 0$
for any $l\in \llbracket{1,n}\rrbracket,$ then $\Sigma _{l=1}^{n}p_{l}H_{l}\in
\mathbf{G}_{n}.$

An $n\times n$ matrix $H$ is said to be a circular matrix if
\begin{equation}
H_{j,j+l}=H_{k,k+l} \ \text{ for all }\{j,k,l\}\subseteq\mathbb{Z}.     \label{eq2-12}
\end{equation}
We define
\begin{eqnarray}
\mathbf{G}_{n,c} &=&\{H:H\in \mathbf{G}_{n}, H \text{ is circular}\},   \label{eq2-13}  \\
\mathcal{B}_{n,c} &=&\{\Pi _{j=1}^{n}H_{j,j+1}:H\in \mathbf{G}_{n,c}\}.     \label{eq2-14}
\end{eqnarray}
By these definitions, one has
\begin{equation}
\mathbf{G}_{n,c}\subseteq \mathbf{G}_{n}, \ \mathcal{B}_{n,c}\subseteq
\mathcal{B}_{n}.     \label{eq2-15}
\end{equation}

For the tuple $\Phi =\left( |\varphi _{1}\rangle ,|\varphi _{2}\rangle
,...,|\varphi _{n}\rangle \right) \in \mathbf{P}_{d}^{n},$ if the Gram
matrix of $\Phi $ is circular, i.e., $G_{\Phi }\in \mathbf{G}_{n,c},$ then
we say that the tuple $\Phi =\left( |\varphi _{1}\rangle ,|\varphi
_{2}\rangle ,...,|\varphi _{n}\rangle \right) $ has the symmetry of circular
Gram matrix. The symmetry of circular Gram matrix for the tuple $\Phi
=\left( |\varphi _{1}\rangle ,|\varphi _{2}\rangle ,...,|\varphi _{n}\rangle
\right) $ implies
\begin{equation}
\langle \varphi _{j}|\varphi _{j+l}\rangle =\langle \varphi _{k}|\varphi
_{k+l}\rangle , \ \{j,k,l\}\subseteq \llbracket{1,n}\rrbracket.     \label{eq2-16}
\end{equation}

Similar to Lemma 2, we have Lemma 3 below.

\begin{Lemma} \label{Lemma-3}  \text{ \ }   \hypertarget{3.a}{}

(3.a). $\mathbf{G}_{n,c}$ is closed under Hadamard product.  \hypertarget{3.b}{}

(3.b). $\mathbf{G}_{n,c}$ is closed under convex combination.
\end{Lemma}

We also need Lemma 4 and Lemma 5 below.

\begin{Lemma} \label{Lemma-4}  \text{ (Theorem 2 in Ref. \cite{Zhang-2025-PRA}). }
For $n\geq 3$, $\mathcal{B}_{n,c}$ is convex.
\end{Lemma}

\begin{Lemma} \label{Lemma-5}  \text{ \ }  \hypertarget{5.a}{}

(5.a). $\mathcal{B}_{n,d}$ is closed under multiplication of complex numbers.     \hypertarget{5.b}{}

(5.b). $\mathcal{B}_{n,c}$ is closed under multiplication of complex numbers.
\end{Lemma}

(5.a) is a result of (2.a), see Proposition 1 in Ref. \cite{Li-2025-PRA}, that is, if $z_{1}\in \mathcal{B}_{n},$  $z_{2}\in \mathcal{B}_{n},$ then  $z_{1}z_{2}\in \mathcal{B}_{n}.$ (5.b) is a result of (3.a).

\subsection{Some existing results about $\mathcal{B}_{n,d}$ and $\boldsymbol{B}%
_{n,d}$}

\bigskip In this subsection, we review some known results for the regions of $%
\mathcal{B}_{n,d}$ and $\boldsymbol{B}_{n,d}.$

For $n=3,$ Fernandes et. al revealed that \cite{Fernandes-2024-PRL}
\begin{equation}
\mathcal{B}_{3}=\{re^{i\theta }\in
\mathbb{C}:1-3r^{\frac{2}{3}}+2r\cos \theta \geq 0\}     \label{eq2-17}
\end{equation}
with $r\geq 0,$ $\theta \in \lbrack 0,2\pi );$ Li et. al proved \cite{Li-2025-PRA} $\mathcal{B}%
_{3}=\mathcal{B}_{3,c}.$

For $n=4,$ Fernandes et. al \cite{Fernandes-2024-PRL} revealed that $\mathcal{B}_{4,c}$ is the region
enclosed by its boundary
\begin{equation}
\partial \mathcal{B}_{4,c}=\left\{ \frac{e^{i\theta }}{\left( \sin \frac{%
\theta }{4}+\cos \frac{\theta }{4}\right) ^{4}}:\theta \in \lbrack 0,2\pi
)\right\} ,    \label{eq2-18}
\end{equation}
and conjectured $\mathcal{B}_{4}=\mathcal{B}_{4,c};$ Zhang et. al \cite{Zhang-2025-PRA} proved $%
\mathcal{B}_{4}=\mathcal{B}_{4,c}.$

For $n\geq 3,$ Li et. al  \cite{Li-2025-PRA} proved that
\begin{equation}
\mathcal{B}_{n,c}=\left\{ t[p+(1-p)\xi _{n}]^{n}:\{t,p\}\subseteq \lbrack
0,1]\right\} ,     \label{eq2-19}
\end{equation}
or equivalently,
\begin{equation}
\partial \mathcal{B}_{n,c}=\left\{ [p+(1-p)\xi _{n}]^{n}:p\in \lbrack
0,1]\right\} ,     \label{eq2-20}
\end{equation}
where $\xi _{n}=e^{i\frac{2\pi }{n}}.$

For $n\geq 3,$ using the
Oszmaniec-Brod-Galv\~{a}o states \cite{Oszmaniec-2024-NJP}
\begin{equation}
|\psi _{k}\rangle =\sin \varphi |0\rangle +\xi _{n}^{k}\cos \varphi
|1\rangle ,\varphi \in \left[ 0,\frac{\pi }{2}\right], k\in \llbracket{1,n}\rrbracket,   \label{eq2-21}
\end{equation}
Li et. al  \cite{Li-2025-PRA} proved that
\begin{equation}
\mathcal{B}_{n,c}\subseteq \mathcal{B}_{n,2}.     \label{eq2-22}
\end{equation}
where $\{|0\rangle ,|1\rangle \}$ is an orthonormal basis of $\mathbb{C}^{2}.$

For $n\geq 3,$ using the Oszmaniec-Brod-Galv\~{a}o states \cite{Oszmaniec-2024-NJP}, Zhang
et. al \cite{Zhang-2025-PRA} proved that Eq. (\ref{eq2-20}) can be rephrased as
\begin{equation}
\partial \mathcal{B}_{n,c}=\left\{ e^{i\theta }\cos ^{n}\frac{\pi }{n}\sec
^{n}\frac{\pi -\theta }{n}:\theta \in \lbrack 0,2\pi )\right\} ,    \label{eq2-23}
\end{equation}
and conjectured $\partial \mathcal{B}_{n}=\partial \mathcal{B}_{n,c}.$

\hypertarget{section III}{}
\section{Regions of $\mathcal{B}_{n,d}$ and $\boldsymbol{B}_{n,d}$}
%\setcounter{equation}{0} \renewcommand%
%\theequation{3.\arabic{equation}}

In this section, we rigorously determine the regions of $\mathcal{B}%
_{n,d}$ and $\boldsymbol{B}_{n,d}.$ The main results of this paper are concluded
in Theorem 1 below.
\begin{Theorem} \label{Theorem-1}
For $n\geq 2$, $d\geq 2,$  it holds that
\begin{equation}
\boldsymbol{B}_{n,d}=\mathcal{B}_{n,d}=\boldsymbol{B}_{n,2}=\mathcal{B}_{n,2}=\mathcal{B}_{n,c}.     \label{eq3-1}
\end{equation}
\end{Theorem}

To prove Theorem 1, we first prove Theorem 2 and Theorem 3.
\begin{Theorem} \label{Theorem-2}
For the tuple of $n$ ($n\geq 2$) pure states $\Psi =\left( |\psi _{1}\rangle
,|\psi _{2}\rangle ,...,|\psi _{n}\rangle \right) $ with $\left\vert
\Pi _{j=1}^{n}\langle \psi _{j}|\psi _{j+1}\rangle \right\vert >0$, there exists a tuple of
$n$ pure states $\Phi =\left( |\varphi _{1}\rangle ,|\varphi _{2}\rangle
,...,|\varphi _{n}\rangle \right) $ such that
\begin{eqnarray}
\langle \varphi _{j}|\varphi _{j+1}\rangle  &=&\langle \varphi _{k}|\varphi
_{k+1}\rangle ,\text{ }\{j,k\}\subseteq \llbracket{1,n}\rrbracket;    \label{eq3-2}   \\
\text{Arg}\left( \Pi _{j=1}^{n}\langle \psi _{j}|\psi _{j+1}\rangle \right)
&=&\text{Arg}\left( \Pi _{j=1}^{n}\langle \varphi _{j}|\varphi _{j+1}\rangle
\right) ;    \label{eq3-3}  \\
\left\vert \Pi _{j=1}^{n}\langle \psi _{j}|\psi _{j+1}\rangle \right\vert
&\leq &\left\vert \Pi _{j=1}^{n}\langle \varphi _{j}|\varphi _{j+1}\rangle
\right\vert,     \label{eq3-4}
\end{eqnarray}
where Arg$(z)$ is the the principal argument of complex number $z,$ i.e., if
we write $z$ in the polar form $z=re^{i\theta }$ with $r\geq 0$ and $\theta
\in \lbrack 0,2\pi ),$ then Arg$(z)=\theta.$
\end{Theorem}

\emph{Proof of Theorem 2.} For $\Psi =\left( |\psi _{1}\rangle ,|\psi _{2}\rangle
,...,|\psi _{n}\rangle \right) $ with $\left\vert
\Pi _{j=1}^{n}\langle \psi _{j}|\psi _{j+1}\rangle \right\vert >0,$ let
\begin{eqnarray}
\langle \psi _{j}|\psi _{j+1}\rangle  &=&r_{j}e^{i\theta _{j}},   \label{eq3-5}  \\
\Pi _{j=1}^{n}\langle \psi _{j}|\psi _{j+1}\rangle  &=&re^{i\theta },    \label{eq3-6}
\end{eqnarray}
with $\{r_{j}\}_{j=1}^{n}\subseteq
\mathbb{R}^{+},$ $r=\Pi _{j=1}^{n}r_{j}>0,$ $\theta \in
\mathbb{R},$ $\{\theta _{j}\}_{j=1}^{n}\subseteq
\mathbb{R}.$ Let $\Phi _{1}=\left( |\psi _{1}^{\prime }\rangle ,|\psi _{2}^{\prime
}\rangle ,...,|\psi _{n}^{\prime }\rangle \right) $ with
\begin{eqnarray}
|\psi _{j}^{\prime }\rangle &=&e^{i\alpha _{j}}|\psi _{j}\rangle , \
j\in \llbracket{1,n}\rrbracket,\ \{\alpha _{j}\}_{j=1}^{n}\subseteq\mathbb{R};     \label{eq3-7}  \\
\alpha _{j} &=&\alpha _{1}+(j-1)\frac{\theta }{n}-\sum_{k=1}^{j-1}\theta
_{j},\ j\in \llbracket{2,n}\rrbracket.    \label{eq3-8}
\end{eqnarray}
One can check that
\begin{equation}
\langle \psi _{j}^{\prime }|\psi _{j+1}^{\prime }\rangle =r_{j}e^{i\frac{%
\theta }{n}},\ j\in \llbracket{1,n}\rrbracket.    \label{eq3-9}
\end{equation}

For any $j\in \llbracket{1,n}\rrbracket$, we define
\begin{eqnarray}
\Psi _{j}=\left( |\psi _{j}^{\prime }\rangle ,|\psi _{j+1}^{\prime }\rangle
,...,|\psi _{n-1}^{\prime }\rangle ,|\psi _{n}^{\prime }\rangle ,|\psi
_{1}^{\prime }\rangle ,|\psi _{2}^{\prime }\rangle ,...,|\psi _{j-1}^{\prime
}\rangle \right). \ \ \ \ \ \   \label{eq3-10}
\end{eqnarray}
From \hyperlink{2.b}{(2.b)} we see that the matrix $\frac{1}{n}\sum_{j=1}^{n}G_{\Psi
_{j}}\in \mathbf{G}_{n}.$ Also, Eqs. (\ref{eq3-9},\ref{eq3-10}) imply
\begin{equation}
\left( \frac{1}{n}\sum_{j=1}^{n}G_{\Psi _{j}}\right) _{k,k+1}=\left( \frac{1%
}{n}\sum_{j=1}^{n}r_{j}\right) e^{i\frac{\theta }{n}}.    \label{eq3-11}
\end{equation}
Lemma \ref{Lemma-1} yields that $\frac{1}{n}\sum_{j=1}^{n}G_{\Psi _{j}}$ corresponds to
a tuple of $n$ pure states $\Phi =\left( |\varphi _{1}\rangle ,|\varphi
_{2}\rangle ,...,|\varphi _{n}\rangle \right) $ such that
\begin{equation}
\langle \varphi _{k}|\varphi _{k+1}\rangle =\left( \frac{1}{n}%
\sum_{j=1}^{n}r_{j}\right) e^{i\frac{\theta }{n}}, \ k\in \llbracket{1,n}\rrbracket.      \label{eq3-12}
\end{equation}
By the inequality of arithmetic mean and geometric mean, we have
\begin{equation}
r=\Pi _{j=1}^{n}r_{j}\leq \left( \frac{1}{n}\sum_{j=1}^{n}r_{j}\right)^{n}.  \label{eq3-13}
\end{equation}
Theorem 2 then follows.
$\hfill\blacksquare$

\begin{Theorem} \label{Theorem-3}
For the tuple of $n$ ($n\geq 2$) pure states $\Phi =\left( |\varphi
_{1}\rangle ,|\varphi _{2}\rangle ,...,|\varphi _{n}\rangle \right) ,$ if
\begin{equation}
\langle \varphi _{j}|\varphi _{j+1}\rangle =\langle \varphi _{k}|\varphi
_{k+1}\rangle , \ \{j,k\}\subseteq \llbracket{1,n}\rrbracket,    \label{eq3-14}
\end{equation}
then there exists a tuple of $n$ pure states $\Omega =\left( |\phi
_{1}\rangle ,|\phi _{2}\rangle ,...,|\phi _{n}\rangle \right) $ such that
\begin{equation}
\langle \phi _{j}|\phi _{j+1}\rangle =\langle \varphi _{j}|\varphi
_{j+1}\rangle , \ j\in \llbracket{1,n}\rrbracket,    \label{eq3-15}
\end{equation}
and also, the Gram matrix of $\Omega ,$ $G_{\Omega },$ is a circular matrix,
i.e., $G_{\Omega }\in \mathbf{G}_{n,c}.$
\end{Theorem}

\emph{Proof of Theorem 3.} For any $j\in \llbracket{1,n}\rrbracket$, we define
\begin{equation}
\Phi _{j}=\left( |\varphi _{j}\rangle ,|\varphi _{j+1}\rangle ,...,|\varphi
_{n-1}\rangle ,|\varphi _{n}\rangle ,|\varphi _{1}\rangle ,|\varphi
_{2}\rangle ,...,|\varphi _{j-1}\rangle \right).     \label{eq3-16}
\end{equation}
From \hyperlink{2.b}{(2.b)} and Eq. (\ref{eq3-16}), we see that the matrix $\frac{1}{n}%
\sum_{j=1}^{n}G_{\Phi _{j}}\in \mathbf{G}_{n,c}.$ Also, Eq. (\ref{eq3-14},\ref{eq3-16}) imply that
\begin{equation}
\left( \frac{1}{n}\sum_{j=1}^{n}G_{\Phi _{j}}\right) _{k,k+1}=\langle
\varphi _{k}|\varphi _{k+1}\rangle , \ k\in \llbracket{1,n}\rrbracket.    \label{eq3-17}
\end{equation}
Lemma \ref{Lemma-1} yields that $\frac{1}{n}\sum_{j=1}^{n}G_{\Phi _{j}}$ corresponds to
a tuple of $n$ pure states $\Omega =\left( |\phi _{1}\rangle ,|\phi
_{2}\rangle ,...,|\phi _{n}\rangle \right) $ such that $\frac{1}{n}%
\sum_{j=1}^{n}G_{\Phi _{j}}=G_{\Omega }.$
Theorem 3 then follows.
$\hfill\blacksquare$

Now we prove Theorem 1.

\emph{Proof of Theorem 1.}
We set this proof into four steps.

(i). The case of $n=2.$

 By the definition of $\mathcal{B}_{2,c}$ in Eq. (\ref{eq2-14}), suppose $z\in\mathcal{B}_{2,c},$ then there exist pure states $\Omega =\left( |\phi _{1}\rangle ,|\phi
_{2}\rangle\right) $ such that $\langle\phi _{1}|\phi _{2}\rangle=\langle\phi _{2}|\phi _{1}\rangle$ and $\langle\phi _{1}|\phi _{2}\rangle \langle\phi _{2}|\phi _{1}\rangle=z$. Let  $|\phi _{1}\rangle=|0\rangle,$ $|\phi _{2}\rangle=\cos\theta|0\rangle+\sin\theta|1\rangle,$ $\theta\in[0,\pi]$, $\{|0\rangle,|1\rangle\}$ is an orthonormal basis of $\mathbb{C}^{2}$, then $\langle\phi _{1}|\phi _{2}\rangle=\langle\phi _{2}|\phi _{1}\rangle=\cos\theta$, $\langle\phi _{1}|\phi _{2}\rangle \langle\phi _{2}|\phi _{1}\rangle=\cos^{2}\theta$. This shows $\{\cos^{2}\theta:\theta\in[0,\pi]\}=[0,1]\subseteq\mathcal{B}_{2,c}$. Obviously, $\mathcal{B}_{2}=\boldsymbol{B}_{2}=[0,1],$  with the fact $\mathcal{B}_{n,c}\subseteq \mathcal{B}_{n}$ in Eq. (\ref{eq2-15}), it follows that $\mathcal{B}_{2,c}=[0,1].$ Combining Eq. (\ref{eq2-9}), we ensure that Eq. (\ref{eq3-1}) holds for $n=2.$

(ii). We prove $\mathcal{B}_{n}=\mathcal{B}_{n,c}$ for $n \geq 3.$

Since $\mathcal{B}_{n,c}\subseteq \mathcal{B}_{n}$ in Eq. (\ref{eq2-15}), we
only need to prove $\mathcal{B}_{n}\subseteq \mathcal{B}_{n,c}.$ From Eq. (\ref{eq2-19}), we see that $0\in \mathcal{B}_{n,c}$. Suppose $0\neq z\in
\mathcal{B}_{n},$ then their exists a tuple of $n$ pure states $\Psi =\left(
|\psi _{1}\rangle ,|\psi _{2}\rangle ,...,|\psi _{n}\rangle \right) $ such
that $\Pi _{j=1}^{n}\langle \psi _{j}|\psi _{j+1}\rangle =z.$ By Theorem \ref{Theorem-2}
and Theorem \ref{Theorem-3}, there exists a tuple of $n$ pure states $\Omega =\left( |\phi
_{1}\rangle ,|\phi _{2}\rangle ,...,|\phi _{n}\rangle \right) $ such that
\begin{eqnarray}
\text{Arg}\left( z\right)  &=&\text{Arg}\left( \Pi _{j=1}^{n}\langle \phi
_{j}|\phi _{j+1}\rangle \right) ;      \label{eq3-18}  \\
\left\vert z\right\vert  &\leq &\left\vert \Pi _{j=1}^{n}\langle \phi
_{j}|\phi _{j+1}\rangle \right\vert,     \label{eq3-19}
\end{eqnarray}
and also, $G_{\Omega }\in \mathbf{G}_{n,c}.$ Since $\mathcal{B}_{n,c}$ is
convex (see Lemma \ref{Lemma-4}) and $0\in \mathcal{B}_{n,c},$
thus there exists $t\in [0,1]$, $z=t\Pi _{j=1}^{n}\langle \phi
_{j}|\phi _{j+1}\rangle\in \mathcal{B}_{n,c}.$ This proves $\mathcal{B}_{n}=\mathcal{B}%
_{n,c}.$

(iii). We prove $\boldsymbol{B}_{n}=\mathcal{B}_{n}$ for $n \geq 3.$

Since $\mathcal{B}_{n}=\mathcal{B}_{n,c}$ and $\mathcal{B}_{n}$ is a convex
and closed set, thus $\mathcal{B}_{n}=$conv$(\mathcal{B}_{n}).$ Together
with the fact $\mathcal{B}_{n}\subseteq \boldsymbol{B}_{n}\subseteq $conv$(%
\mathcal{B}_{n}),$ it follows that $\boldsymbol{B}_{n}=\mathcal{B}_{n}.$

(iv). We prove $\boldsymbol{B}_{n}=\boldsymbol{B}_{n,2}$ for $n \geq 3.$

$\boldsymbol{B}_{n}=\boldsymbol{B}_{n,2}$ comes from the facts $\boldsymbol{B%
}_{n,2}\boldsymbol{\subseteq B}_{n}$ in Eq. (\ref{eq2-5}) and $\mathcal{B}_{n,c}\boldsymbol{%
\subseteq B}_{n,2}$ in Eq. (\ref{eq2-22})

We then completed the proof of Theorem 1.
$\hfill\blacksquare$

Note that, in Theorem \ref{Theorem-1} we suppose $n\geq 2$. If $n=1,$ the following Proposition 1 obviously holds.

\begin{Proposition} \label{Proposition-1}
For $n=1$, $d_{1}\geq 1,$  $d_{2}\geq 1,$ it holds that
\begin{equation}
\boldsymbol{B}_{1,d_{1}}=\mathcal{B}_{1,d_{1}}=\boldsymbol{B}_{1,d_{2}}=\mathcal{B}_{1,d_{2}}=\mathcal{B}_{1,c}=\{1\}.     \label{eq3-20}
\end{equation}
\end{Proposition}

Combining Lemma \ref{Lemma-4}, Lemma \ref{Lemma-5}, and Theorem \ref{Theorem-1}, we see that $\mathcal{B}_{n,d}$ and $\boldsymbol{B}_{n,d}$ are convex and closed under multiplication of complex numbers.

\hypertarget{section IV}{}
\section{Summary}
In this work, we have completely characterized the numerical ranges of Bargmann invariants for both pure and mixed states in finite-dimensional quantum systems. Our analysis reveals a remarkable simplification: all physically realizable values of these invariants can be attained using either pure states with circular Gram matrix symmetry or general qubit states. This result demonstrates that the complex behavior of Bargmann invariants in high-dimensional systems is fundamentally constrained by these two special cases.

Given the crucial role of Bargmann invariants in quantum information science, our complete characterization provides essential theoretical tools for these applications. The explicit determination of these numerical ranges not only resolves a long-standing open question in quantum mechanics but also opens new possibilities for utilizing Bargmann invariants in quantum metrology, error mitigation protocols, and quantum resource theories. These findings are expected to significantly advance both theoretical and experimental studies of quantum systems.
\section*{ACKNOWLEDGMENTS}

This work was supported by the National Natural Science Foundation of China (Grant No. 12471443).

%\bibliographystyle{apsrev4-1}
%\bibliography{Bargmann-invariants}

%

\end{document}